\documentclass[prl,twocolumn,english,longbibliography,superscriptaddress]{revtex4-1}

\usepackage{amsmath}
\usepackage{amssymb}
\usepackage{amsthm}

\usepackage{amsbsy}
\usepackage{amstext}
\usepackage{graphicx}
\usepackage{color}
\usepackage{float}

\usepackage{wasysym}

\makeatletter
\usepackage{amsfonts}
\usepackage{dcolumn}
\usepackage{bbold,bm}

\def\re{\mathrm{Re}\,}
\def\im{\mathrm{Im}\,}

\def\id{\mathbb{1}} % need bbold,bm package

\def\Texp{\mathbb{T}\exp}

\def\nodag{^{\vphantom{\dagger}}}
\def\floor#1{\lfloor{#1}\rfloor}
\def\mod{\:\mathrm{mod}\:}
\def\trans{{\raisebox{-1pt}{{\scriptsize\textsf{T}}}}}
\def\zc#1{{#1}^\natural}        % --- Floquet zone center
\def\zcsym{\natural}        % --- Floquet zone center symbol
\def\ze#1{{#1}^\flat}   % --- Floquet zone edge
\def\zesym{\flat}   % --- Floquet zone edge symbol
\def\zezc#1{{#1}^\circ} % --- Floquet zone edge OR Floquet zone center
\def\zezcsym{\circ} % --- Floquet zone edge OR Floquet zone center symbol

\newcommand{\beq}    {\begin{equation}}
\newcommand{\enq}    {\end{equation}}

\makeatother

\begin{document}

\title{Universal fluctuations of Floquet topological invariants at low frequencies}

\author{M. Rodriguez-Vega}
\affiliation{Department of Physics, Indiana University, Bloomington, Indiana 47405,
USA}

\author{B. Seradjeh}
\affiliation{Department of Physics, Indiana University, Bloomington, Indiana 47405,
USA}
\affiliation{Max Planck Institute for the Physics of Complex Systems, N\"othnitzer Str. 38, 01187 Dresden Germany}

\begin{abstract}
We study the low-frequency dynamics of periodically driven one-dimensional systems hosting Floquet topological phases. We show, both analytically and numerically, in the low frequency limit $\Omega\to0$, the topological invariants of a chirally-symmetric driven system exhibit universal fluctuations. While the topological invariants in this limit nearly vanish on average over a small range of frequencies, we find that they follow a universal Gaussian distribution with a width that scales as $1/\sqrt{\Omega}$. We explain this scaling based on a diffusive structure of the winding numbers of the Floquet-Bloch evolution operator at low frequency. We also find that the maximum quasienergy gap remains finite and scales as $\Omega^2$. Thus, we argue that the adiabatic limit of a Floquet topological insulator is highly structured, with universal fluctuations persisting down to very low frequencies. 
\end{abstract}

%\date{\today}

\maketitle

% Introduction---
% Floquet topology
% Low frequency; adiabatic limit
% Previous work -- high winding numbers -- insight
% Summary of main results; significance

The behavior of a periodically driven system can be qualitatively different from its equilibrium behavior. Manifestations of such behavior in classical physics include resonances, dynamical stabilization of new steady states, and the period-doubling approach to chaos~\cite{kapitza1951,Fei78a,ref1d}. In quantum systems, the effective Floquet dynamics of a driven systems has been employed as a powerful way to engineer designer Hamiltonians, e.g. by using laser sequences in cold atomic gases. In this way, novel phases of matter have been proposed and realized~\cite{JotMesDes14a,AidLohSch15a,zhang2017x,choi2017x,Eck17a}. 

More recently, it has been understood that a driven system can also exhibit essentially non-equilibrium \emph{topological} phases, dubbed Floquet topological phases  \cite{oka2009,kitagawa2011,lindner2011,jiang2011}. Drive parameters, such as the frequency $\Omega$ or the shape of the drive (``drive protocol'') have been proposed~\cite{zhenghao2011,dora2012,KunSer13a,
KunFerSer14a,perez2014,
ref1a,ref1c,
titum2015,
KunFerSer16a}
and used in the lab~\cite{rechtsman2013,WanSteJar13a,
SieLuiLee17a,ChePanWan18a}
to engineer a rich array of topological phases not possible in equilibrium systems. The non-equilibrium dynamics at large frequencies is relatively well understood, e.g. within rotating-wave approximation, as a renormalization of the equilibrium parameters of the system~\cite{blanes2009,rahav2003,rahav2003b,bukov2015,EckAni15a}. The low-frequency regime, on the other hand, remains largely unexplored~\cite{GomPla13a}. This is the relevant regime in solid-state systems driven by ac potentials~\cite{lindner2011}. It is also important as a way to reduce unwanted heating in the system~\cite{LazDasMoe14a,DAlRig14a,WeiKna17a,StrEck16a}. At a more basic level, it relates to the adiabatic limit as $\Omega\to0$. Numerical studies have reported nonzero Floquet topological invariants as frequency is lowered~\cite{KunSer13a,LiuLevBar13a,TonAnGon13a,GomDelPla14a,MikKitYas15a,ref1b}. This raises questions on the nature of adiabatic limit in Floquet topological phases.

In this paper, we study the low-frequency limit of one-dimensional model driven systems that exhibit a rich Floquet topological phase diagram~\cite{jiang2011,KunSer13a}. Assuming the driven systems are chirally symmetric~\cite{asboth2014,ref1b,dallago2015}, we derive analytical expressions for the Floquet topological invariants and evaluate them numerically over several decades of the drive frequency. We find that these topological invariants not only remain nonzero at low frequencies, but increasingly fluctuate. While at any fixed frequency the invariants are deterministic, over a range of frequencies $\delta\Omega \ll \Omega$, the invariants distribute pseudorandomly. We argue that this distribution is universal and in our models is given by a Gaussian, whose width is $\sigma(\Omega) \sim 1/\sqrt{\Omega}$. We explain this universal behavior by revealing a diffusive process in the evaluation of the invariants and confirm our results numerically.

Specifically, we study one-dimensional driven systems with periodic boundary conditions, with a Hamiltonian of the form  $\hat H=\int  \hat c^\dagger_k h\nodag_k \hat c\nodag_k \frac{dk}{2\pi}$, where $k\in[-\pi,\pi]$ is the crystal momentum, $\hat c_k$ is a two-component spinor field, and  $h_k = d_{kx} \sigma_x + d_{ky} \sigma_y$ with $d_{kx} + i d_{ky} \equiv d_k$ a model-dependent function. For example, in the Su-Schrieffer-Heeger (SSH) model~\cite{su1979,supp}
$
d_k = 2e^{ik/2} \left(w \cos \frac k2 + i\delta \sin \frac k2 \right),% = |d_k| e^{i(\beta_k+k/2)},
$
where $w$ ($\delta$) is the hopping (modulation) amplitude. In the Kitaev model~\cite{Kit01a,supp}, after a suitable rotation in the Nambu space, one finds $d_k = 2w\cos k - \mu + i\Delta\sin k$, where $\mu$ is the chemical potential and $\Delta$ is the nearest-neighbor pairing amplitude.

These Hamiltonians are particle-hole symmetric, $\sigma_zh_{-k}^*\sigma_z = -h_k$, with eigenvalues $\pm |d_k|$. In equilibrium, there are two topologically distinct phases: a topological phase, for $\delta/w>0$ in the SSH and $|\mu|<2|w|$ for Kitaev model, and a trivial phase otherwise. These two phases are distinguished on the lattice with open boundary conditions by the presence of zero-energy bound states in the topological phase. With periodic boundary conditions, the phases are distinguished by an integer topological invariant $\nu=0$ or $1$, equal to the winding number %of the periodic complex function $d_k$, 
%%%%
\begin{equation}
\nu = \mathcal{W}[d] \equiv \frac1{2\pi i} \int_{-\pi}^\pi \frac{\partial}{\partial k}\ln (d_k) dk.
\end{equation}
% = (\pi+\beta_\pi-\beta_{-\pi})/{2\pi}$. %\frac12+\frac{\beta_\pi-\beta_{-\pi}}{2\pi}
For a multi-band system, e.g. the SSH-Kitaev~\cite{WatEzaTan14a,supp}, $d_k$ is matrix-valued and the topological invariant is found by $\mathcal{W}[\det d]$.

When the system is periodically driven, the full dynamics is obtained by solving the Floquet-Schr\"odinger equation $[h_k(t)-i\partial_t] \phi^\pm_k(t) = \pm\epsilon_k \phi^\pm_k(t)$ (we are setting $\hbar=1$) for the periodic steady states $\phi^s_k(t)=\phi^s_k(t+2\pi/\Omega)$, $s=\pm$, with the quasienergy $s\epsilon_k$, which we take to be in the Floquet zone $[-\Omega/2,\Omega/2]$. The Bloch evolution operator can then be written as $U_k(t) = \sum_{s=\pm} e^{-is\epsilon_k t} \phi_k^s(t) \phi_k^{s\dagger}(0)$. The full-period evolution operator $U_k(2\pi/\Omega)$ has eigenstates $\phi^s_k(0)$ with eigenvalues $e^{-2s\pi i\epsilon_k/\Omega}$. Since the quasienergy is a modular quantity, even a two-band model is characterized by two gaps at Floquet zone center ($\zc\epsilon=0$) and Floquet zone edge ($\ze\epsilon=\Omega/2$)~\cite{Note3}. Thus, for periodic boundary conditions there are \emph{two} independent topological invariants defined for the quasienergy gaps at Floquet zone center, $\zc\nu$, and edge, $\ze\nu$. For open boundary conditions, the corresponding invariants are the number of midgap steady bound states at Floquet zone center and edge~\cite{supp}.

%%%%% Fig. 1-old %%%%%%
%\begin{figure}[tb]
%	\centering
%	\includegraphics[width=3.3in]{fig1.pdf}
%	\caption{Quasienergy spectrum of the driven SSH model with open boundary conditions as a function of frequency. The chain has $1000$ sites and is subject to a twos-step drive protocol switching between $\delta_1/w = -0.75$ and $\delta_2/w = -0.25$. Edge modes within quasienergy gaps at $\zc\epsilon=0$ and $\ze\epsilon=\Omega/2$ are found in an increasing number as the frequency is lowered.
%	}
%	\label{fig:spectrum_critical}
%\end{figure}

To simplify our discussion, we take the drive protocol to satisfy the chiral reflection symmetry, $\delta(t+\pi/\Omega)=\delta(-t+\pi/\Omega)$; then, the two topological invariants are found~\cite{asboth2014,supp} from the half-period evolution operator
%%%%
%\begin{equation}
$
U_k(\pi/\Omega) \equiv \left(\begin{array}{cc} A_k & B_k \\ C_k & D_k  \end{array}\right),
$
%\end{equation}
%%%%
as
%%%%
\begin{equation}
\zc\nu = \mathcal{W}[B]\quad\text{and}\quad\ze{\nu}=\mathcal{W}[D].
\end{equation}
%%%%
In the static case, $D_k$ is constant and $B_k \propto d_k$, thus one finds $\ze\nu = 0$ and $\zc\nu = \nu$ as expected. For concreteness, we present our results for the SSH model in the following and for other models in the Supplemental Material~\cite{supp}.

At symmetry points $k_s=0, \pm\pi$, $h_{k_s}\propto\sigma_x$ and the half-period evolution operator takes simple forms,
%%%%
%\begin{align}
$U_0(\pi/\Omega) = e^{-i\frac{2\pi}{\Omega} w\sigma_x}$
and
$
U_\pi(\pi/\Omega) = e^{i\frac{2\pi}{\Omega}\bar\delta\sigma_x},
$
%\end{align}
%%%%
where $\bar\delta=(\Omega/2\pi)\int_0^{2\pi/\Omega}\delta(s)ds$ is the average hopping modulation through one drive cycle. The values $D_{\pm\pi} = \cos(2\pi\bar\delta/\Omega)$, $D_0 = \cos(2\pi w/\Omega)$ and $B_{\pm\pi} = i\sin(2\pi\bar\delta/\Omega)$, $B_0 = -i\sin(2\pi w/\Omega)$ can be used to anchor their winding.

\begin{figure}[tb]
	\centering
    \includegraphics[width=2.6in]{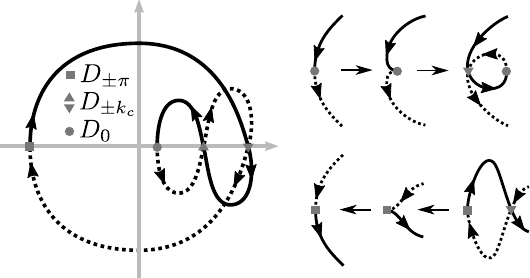}
	\caption{The topological invariant $\ze\nu$ is the winding number of the complex function $D_k$, $k\in[-\pi,\pi)$ (left), computed from the crossing points on the real axis. The inversion symmetry of the SSH model yields a reflection symmetric contour around the real axis, the solid (dashed) line designating the portion corresponding to $k\in[-\pi,0]$ ($[0,\pi]$). As the frequency is lowered, new crossing points are emitted from $D_0$ when the contour twists (top right) and absorbed into $D_{\pm\pi}$ when it untwists (bottom right).
	}
	\label{fig:knot}\vspace{-4mm}
\end{figure}

To understand the changes in the winding number $\ze\nu$ ($\zc\nu$) we analyze the contour of $D_k$ ($B_k$) in the complex plane as frequency varies (see Fig.~\ref{fig:knot}). At high enough frequency the contour of $D_k$ ($B_k$) is a loop with two crossing points on the real (imaginary) axis at $D_{\pm\pi}$ and $D_{0}$ ($B_{\pm\pi}$ and $B_0$); as frequency is lowered the loop twists and untwists, thus changing the number of crossing points on the real (imaginary) axis via two processes: a pair of crossings are ``emitted'' from $D_0$ ($B_0$)  whenever $D'_0 = 0$ ($B'_0 = 0$), where the prime denotes $\partial/\partial k$; on the other hand, a pair of crossings are ``absorbed'' into $D_{\pm\pi}$ ($B_{\pm\pi}$) when $D'_{\pm\pi} = 0$ ($B'_{\pm\pi} = 0$). While the rates of these processes depend on the drive protocol, they all scale with $1/\Omega$; thus the number of crossings generically grows as $1/\Omega$. As $\Omega$ is lowered, all crossings \emph{move} back and forth within the unit disk along the real (imaginary) axis at a speed that scales with $1/\Omega$. When a crossing point of $D_k$ ($B_k$) passes through the origin, the winding $\ze\nu$ ($\zc\nu$) changes. The inversion symmetry of the SSH model ensures that except $D_{\pm\pi}$ and $D_0$ ($B_{\pm\pi}$ and $B_0$), all other crossings are doubled.
%\notemr{Do emittion and absorbtion ocurr and $0$ and $\pm \pi$ for all systems? -- Yes!}

%%%% Fig. 1 %%%%%%

%%%%%% Fig. 2 %%%%%%%
\begin{figure*}[ht]
	\centering
	\includegraphics[width=6.95in]{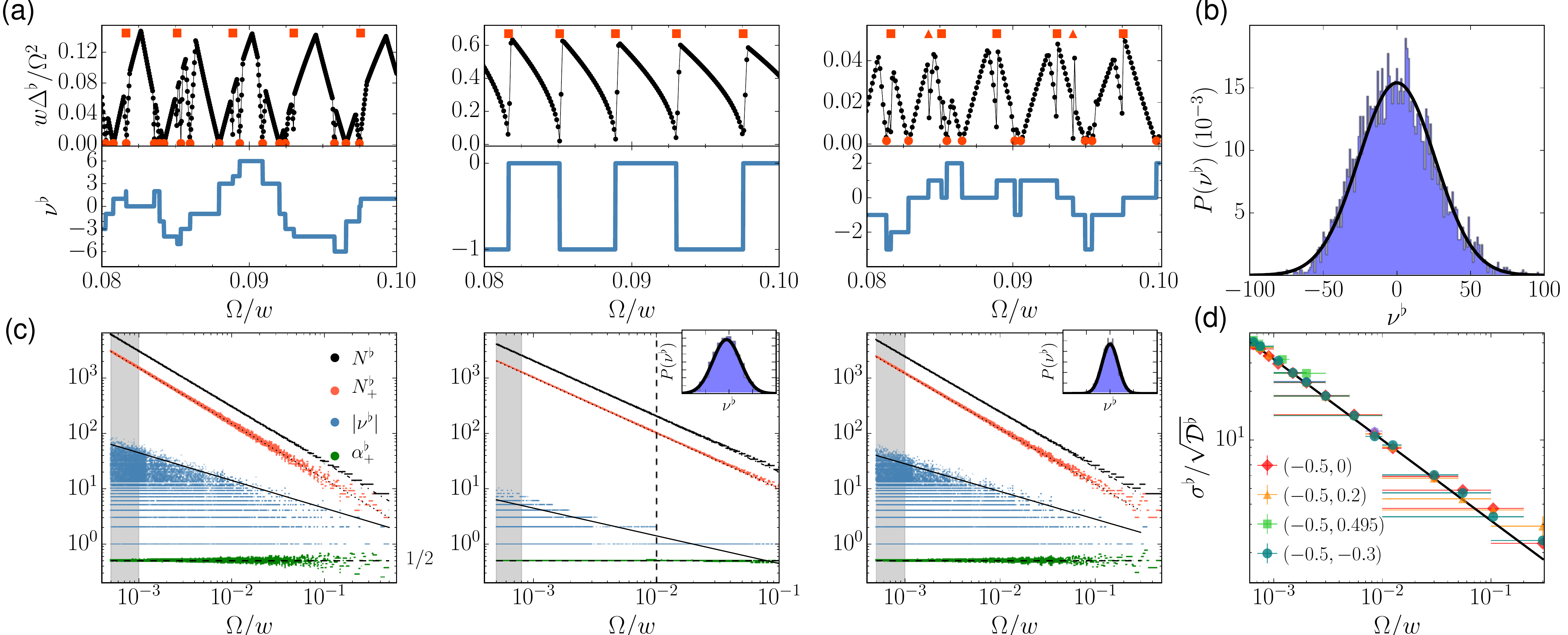}
	\caption{(color online)
	(a) The quasienergy gap (top panels) and the topological invariant (bottom panels) at the Floquet zone edge as a
	function of the frequency. The two-step drive protocols are critical ($\delta_1/w=-0.5, \delta_2/w=0$, left), asymmetric ($\delta_1/w=-0.5, \delta_2/w=0.495$,
	center), and trivial ($\delta_1/w=-0.5, \delta_2/w=-0.3$, right). In all cases $p_1=p_2=0.5$. The (orange) markers indicate analytically calculated gap closings at the symmetric points $k_s=0$ (square), and $k_s=\pm\pi$ (triangle) and non-symmetric points $\ze k_c\neq k_s$ (circle).
	(b) Probability distribution of $\ze\nu$ for the critical protocol in (a) and frequency range $\Omega/w \in (0.5,1)\times10^{-3}$. The solid line is a Gaussian fit.
	(c) Statistics of crossing points and winding number for the three drive protocols in (a). From top to bottom in each panel: the total number of crossing points, $\ze N$, the positively-oriented crossing points $\ze N_+$, the ratio $\ze \alpha_+=\ze N_+/\ze N$, and the winding number $|\ze\nu|$ are calculated numerically at 65000 frequencies. At this resolution, fluctuations in the winding numbers render their graphs random. The vertical dashed line in the center panel marks twice the value of the asymmetry parameter. The insets show the probability distribution $P(\ze\nu)$ over the shaded range as in (b). The solid line shows $\sqrt{\ze{\mathcal{D}}/\Omega}$ with $\ze{\mathcal{D}} = ||\delta_1|-|\delta_2||$.
	(d) Standard deviation $\ze \sigma(\Omega)$ for various two-step drive protocols found by a Gaussian fit to $P(\ze\nu)$. The horizontal (vertical) line at each point indicates the range of frequencies (fitting error). The legend shows the values $(\delta_1/w, \delta_2/w)$. The solid line is $\ze \sigma = \sqrt{\ze{\mathcal{D}}/\Omega}$. 
	}
	\label{fig:main}\vspace{-4mm}
\end{figure*}
%%%%%%%%%%%%%%%%%

Denoting the momenta at crossing points with $\zezc k_c(\Omega)$, where $\zezcsym={\zcsym,\zesym}$, the total number of crossings is $\zezc N=\zezc N_++\zezc N_-$, where 
%%%%%
$\ze N_\pm = \sum_c\Theta(\pm D_{\ze k_c}\, \im D'_{\ze k_c})$ 
%%%%%
and
%%%%%
$\zc N_\pm = \sum_c\Theta(\pm iB_{\zc k_c}\, \re B'_{\ze k_c})$.
%%%%%
The winding numbers, on the other hand, are given by $\zezc\nu=\frac12(\zezc N_+-\zezc N_-)$.  At any given frequency, $\Omega$, the values of $\zezc N_\pm$ may be computed deterministically from the number of crossings emitted, absorbed, and moved on the corresponding real or imaginary axis. However, as $\Omega\to0$, these numbers grow in an increasingly complex way; thus, over a frequency interval $\delta\Omega\ll\Omega$ the distribution of crossing points appears random. We posit that this distribution can be modeled by a universal stochastic process of emission, absorption, and motion of crossing points of $D_k$ ($B_k$)~\cite{Note0}. In the low-frequency limit, our numerics show generically that $\zezc N_\pm$ are equally distributed. Taking this to be true, we may think of $\zezc N_\pm$ as the number of steps taken by a one-dimensional random walker in opposite directions, with $2\zezc\nu$ the distance from the starting point. Thus, winding numbers are diffusive variables with a protocol-dependent diffusion constant $\zezc{\mathcal{D}}= 2\Omega \sqrt{\langle {\zezc N_+}^2\rangle}$. Here, $\langle\cdots\rangle$ stands for the average in the stochastic model or, equivalently, the average over the interval $\delta\Omega$. The winding numbers acquire a Gaussian distribution with a width $\zezc\sigma(\Omega) = \sqrt{\zezc{\mathcal{D}}/\Omega}$. This is our main result.

Changes in the winding number are concomitant with quasienergy gap closings. This is easy to see at symmetry points $k_s$, where, for our chirally symmetric protocols, the full-period evolution operator is the square of the half-period evolution operator. At these points, $\ze\nu$ ($\zc\nu$) change by one when $D_0$ and $D_{\pm\pi}$ ($B_0$ and $B_{\pm\pi}$) vanish, respectively, at $\ze\Omega_{0} = \frac{4w}{2m-1}$ and $\ze\Omega_{\pi}=\frac{4\bar\delta}{2m-1}$ ($\zc\Omega_{0} = \frac{2w}{m}$ and $\zc\Omega_{\pi}=\frac{2\bar\delta}{m}$) for integer $m$. Noting that quasienergies at symmetry points are given by $\epsilon_0\equiv\pm2w \mod\Omega$ and $\epsilon_\pi\equiv\pm2\bar\delta \mod\Omega$, it is easy to see they are equal to $\ze\epsilon$ ($\zc\epsilon$) exactly at frequencies where $\ze\nu$ ($\zc\nu$) changes. Of course, changes in $\ze\nu$ ($\zc\nu$) are also caused at any frequency $\ze\Omega_{*}$ ($\zc\Omega_{*}$) and non-symmetry momenta $\ze k_*\equiv \ze k_c(\ze\Omega_*)$ [$\zc k_*\equiv \zc k_c(\zc\Omega_*)$], where $D_{\ze k_*}$ ($B_{\zc k_*}$) vanishes and the gap at $\ze\epsilon$ ($\zc\epsilon$) closes. Due to inversion symmetry, the winding numbers at these gap closings change by two. We note that  the frequencies $\ze\Omega_*$ and $\zc\Omega_*$ depend on the drive protocol.

%Technical discussion ---
%\notemr{technical discussion} 
%
% Two-step drive
To proceed quantitatively, we choose a periodic two-step drive protocol in the SSH model given by
%$\delta(t) = \delta_1 \left(\Theta(\Omega t) - \Theta(\Omega t- 2\pi p_1) \right) + \delta_2 \left(  \Theta(\Omega t - 2\pi p_1) - \Theta(\Omega t - 2\pi ) \right)$,
$\delta(t) = \delta_1$ for $0<t<2\pi p_1/\Omega$ and $\delta_2$ for $2\pi p_1/\Omega<t<2\pi/\Omega$.
Here, $0<p_1<1$ is the
dimensionless fraction of the period for the first step of the drive. 
This family of protocols simplifies the numerical calculations, and allows us to obtain both
analytically exact and numerically reliable results over a wide range of frequencies. 
Note that the modulation is chiral symmetric. This is explicit if we take the origin of time to be at $\pi p_1/\Omega$. 
%Explicitly, we change basis with the unitary $\exp(- \pi i p_1 h_{1k} /\Omega)$ to obtain $U_k(2\pi/\Omega) =\exp(- 2\pi i p_2 h_{2k}/\Omega)\exp(- 2\pi i p_1 h_{1k}/\Omega)$, where $p_2=1-p_1$ and $h_a = d_{ak} \sigma_x + d_{ay} \sigma_y$ with $d_a = d_{ax} + i d_{ay} = 2e^{ik/2} \left(w \cos \frac k2 + i\delta_a \sin \frac k2 \right)$ for $a=1,2$.**** 
%In terms of the Pauli matrices, the Floquet propagator can be written as $U(T) = \alpha_0 + i (\alpha_x \sigma_x + \alpha_y \sigma_y) $, with $\alpha_i=\alpha_i(k,\Omega) \in \mathbb{R}$. The coefficients $\alpha_i(k,\Omega)$ are also $\delta_1$-, and $\delta_2$-dependent and are not shown explicitly. 
Calculating the full-period evolution operator, the quasienergies are given by
%%%%%
\begin{equation}
\cos \frac{2\pi\epsilon_k}\Omega  = 	\cos \frac{2\pi\bar E_k}\Omega \cos^2 \frac{\theta_k}2 
			+ \cos \frac{2\pi\breve E_k}\Omega \sin^2\frac{\theta_k}2,
\end{equation}
%%%%%
where the average and difference bands $\bar E_k = p_1 |d_{1k}| + p_2 |d_{2k}|$, $\breve E_k = p_1|d_{1k}| - p_2|d_{2k}|$ with index $a=1,2$, indicating $\delta=\delta_a$. Here, $\theta_k$ is the angle between the complex variables $d_{1k}$ and $d_{2k}$. Without loss of generality, we assume $|\delta_1| > |\delta_2|$. Gap closings at $\zezc\epsilon$ for $\zezc k_*\neq 0,\pi$ are obtained when 
$\cos (2\pi \bar E_{\zezc k_*}/\Omega)  = \cos (2\pi \breve E_{\zezc k_*} /\Omega ) = \zezc{(-1)}$, where $\ze{(-1)}=-1$ and $\zc{(-1)} = 1$. 
This is a resonant condition leading to an implicit equation for $\Omega$, which we solve numerically. 
Furthermore, for $|\delta_2/\delta_1| p_2 < p_1 < p_2$,  there exist $\zc k_*$ where $\breve E_{\zc k_*} = 0$; at these points, the gap at $\zc\epsilon$ closes
for  $\bar E_{\zc k_*}/\zc \Omega_*  \equiv 0 \mod 1$.

The winding numbers are found from
%%%%
\begin{equation}
D_k = e^{-i\theta_k/2} \left( \cos\frac{\pi\bar E_k}{\Omega}\cos\frac{\theta_k}2 + i \cos\frac{\pi\breve E_k}{\Omega}\sin\frac{\theta_k}2 \right)
\end{equation}
%%%%
and $B_k=(d_{1k}/|d_{1k}|)\tilde B_k$, %e^{i(\beta_{1k}+k/2)}
%%%%
\begin{equation}
\tilde B_k = e^{-i\theta_k/2} \left( \sin\frac{\pi\bar E_k}{\Omega}\cos\frac{\theta_k}2 + i \sin\frac{\pi\breve E_k}{\Omega}\sin\frac{\theta_k}2 \right).
\end{equation}
%%%%
Since $\theta_\pi = 0$ or $\pi$ and $\theta_0=0$, $D_\pi$ ($\tilde B_\pi$) and $D_0$ ($\tilde B_0$) are real.  We note $\mathcal{W}[B]=\mathcal{W}[d_1]+\mathcal{W}[\tilde B]$, i.e. $\zc\nu=\nu_1+\mathcal{W}[\tilde B]$. 

Focusing on $\ze\nu$ for concreteness and using the explicit forms of $D_k$, we find that the crossing points that contribute to either $N_+$ or $N_-$ are emitted when $2p_1w/\Omega$ or $2p_2w/\Omega$ is an integer, and they are absorbed when $p_1|\delta_1|/\Omega$ or $2p_2|\delta_2|/\Omega$ is an integer. For $p_1\approx p_2$ and for small enough frequency, we may assume the motion of crossing points yields a nearly uniform distribution along the real axis. Since the winding number varies by 2 only when the crossing two points are on different halves of the real axis, the diffusion constant may be obtained by  $\ze{\mathcal{D}} \approx 2|p_1(w-|\delta_1|)-p_2(w-|\delta_2|)|$. In the following, we assume $\delta_1<0$ for simplicity. 

%%%%% Fig. 3 %%%%%%
\begin{figure}[tb]
	\centering
	\includegraphics[width=8.5cm]{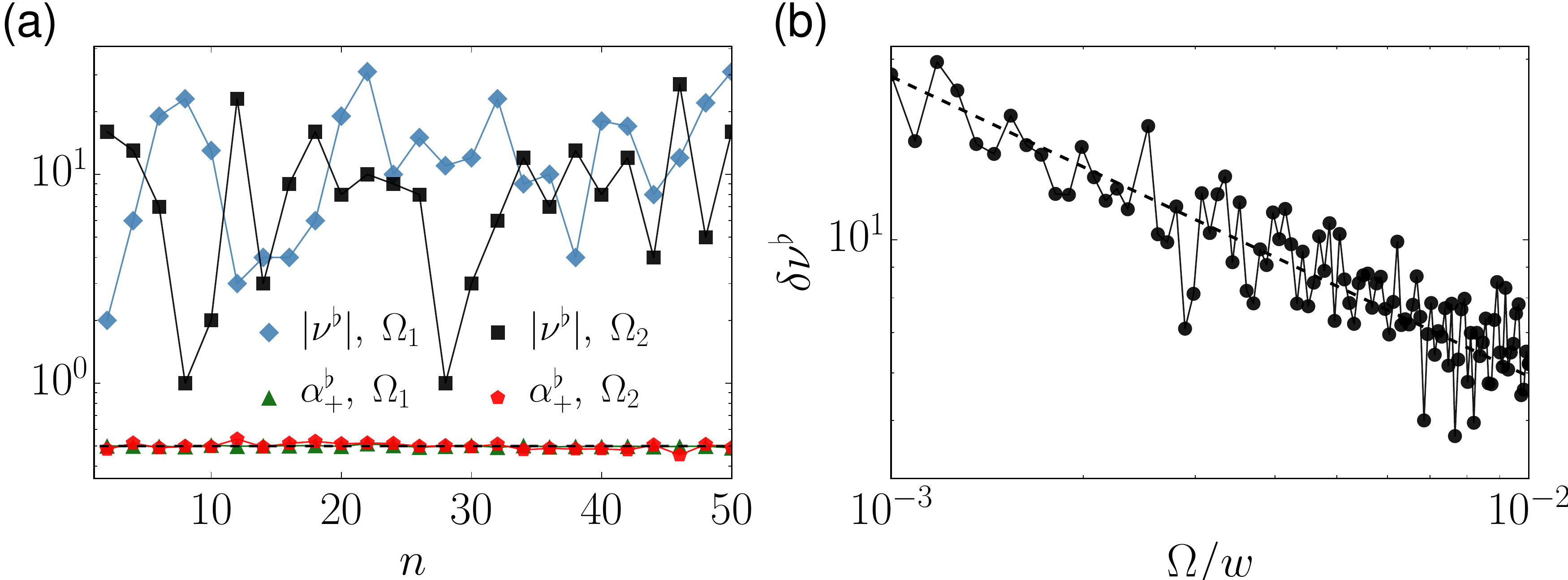}
	\caption{(color online)
	(a) Winding number, $|\ze\nu|$, and the ratio $\ze\alpha_+$ 
		in a multi-step protocol as a function of the number of steps, $n$. The 
		two frequencies are
		$\Omega_1/w = 1.09\times 10^{-3}$ and $\Omega_2/w = 4.15 \times 10^{-3}$, 
		with the mean value
		$\bar\delta/w =-0.4$ and amplitude $A/w = 0.1$.
		(b) The root-mean-square, $\delta\ze\nu$, averaged over $n$ as a function frequency.
		The dashed line corresponds to $\sqrt{\ze{\mathcal{D}}/\Omega}$, fitted with $\ze{\mathcal{D}}/w = 0.35 $.
	}
	\label{fig:nstep}\vspace{-4mm}
\end{figure}
%%%%%%%%%%%%%%%

% three protocols: symmetric, asymmetric, trivial 
Our analytical expressions for the two-step drive allow for the exact determination of gap closings; however, in general, quasienergy gaps and topological invariants can only be obtained by numerical solutions. In the special case of a symmetric drive, $\delta_2=-\delta_1$ and $p_1=p_2$, we can calculate the topological invariants exactly: $\zc\nu = 0$, and $\ze\nu = -1$ when $\lfloor \frac{2w}\Omega-\frac12 \rfloor$ is even and 0 otherwise. 

For the numerical solutions, we consider three distinct protocols: the \emph{asymmetric protocol}, $\delta_2 = - \delta_1 - \varrho > 0$, which is a periodic switch between the equilibrium trivial and the topological phases with the asymmetry parameter $\varrho$; the \emph{critical protocol}, $\delta_2 = 0$, i.e. a periodic switch between the equilibrium trivial and the critical point of the system; and finally, the \emph{trivial protocol}, $\delta_1 < \delta_2 < 0$, such that the systems is in the equilibrium trivial phase at all times. Our numerical results for $\ze\nu$ are summarized in Fig.~\ref{fig:main}; our results for $\zc\nu$ are similar~\cite{supp}.

% quasienergy gaps vs. Omega
The quasienergy gaps $\zezc\Delta= \min_k|\epsilon_k-\zezc\epsilon|$ exhibit self-similar patterns, 
with peaks that scale as $\sim \Omega^2$. %Note that the relative error in the calculation is smaller than the size of the data points. 
We have benchmarked our numerical calculation with the exact analytical expressions for gap closings, shown on the same plot; the agreement is extremely good. The $\Omega^2$ scaling can
be understood within adiabatic perturbation theory~\cite{RigOrtPon08a,RodLenSer18a}, where the frequency is used as a perturbation parameter. The first-order correction to the quasienergy is the Berry phase of the steady states, which vanishes for our chirally symmetric protocols~\cite{RodLenSer18a,MarMoi15a}.
%The first non-zero correction is of order $\Omega^2$. 

% winding numbers
The winding number $\ze\nu$ fluctuates as frequency is lowered with an increasing relative amplitude. For an asymmetric protocol, as in Fig.~\ref{fig:main}(c) center panel, when $\Omega>2\varrho$, we observe the regular step-wise behavior as in the symmetric protocol. However, when $\Omega<2\varrho$, the same fluctuating pattern sets in. We have carried out a detailed analysis of $\ze\nu$ over a wide range of low frequencies.
For each frequency $\Omega$, we show $\ze N$, 
$\ze N_+$, $\ze \alpha_+ = \ze N_+/\ze N$, and $\ze \nu$. Both $\ze N$, and $\ze N_+$ 
scale linearly with $1/\Omega$, confirming our general arguments. The ratio $\ze \alpha_+$ approaches $1/2$ as the frequency 
decreases, indicating the diffusive behavior familiar from a random-walk process. Moreover, the range of $|\ze\nu|$ scales as 
$\sim1/\sqrt{\Omega}$. For a range of frequencies much lower than other energy scales in the system, we have determine the probability $P(\ze\nu)$ of finding 
$\ze\nu$ in our numerical histogram. As shown, $P(\ze \nu)$ follows a Gaussian distribution with a width that is given by $\sigma(\ze\nu) = \sqrt{\mathcal{D}^\flat/\Omega}$ over several decades of frequency, confirming our general result.

% multi-step drive
To test our general arguments for the universality of the fluctuations, we have studied other drive protocols and other models, including a model with multiple bands. These numerical studies support our results in all cases. Details are found in the Supplemental Material~\cite{supp}; here, we present our results for a multi-step protocol in the SSH model approximating $\delta(t) = \bar\delta + A \cos ( \Omega t )$. The analytical calculations become increasingly difficult as the number of steps $n$ in the drive increases; however, we can still calculate the topological invariants numerically. A typical sampling of our results are shown in Fig.~\ref{fig:nstep}. While $\ze\nu$ fluctuates both in magnitude and sign as $n$ is varied, the ratio $\ze\alpha_+\approx1/2$, again indicating a diffusive process. Collecting good statistics over a wide frequency range quickly becomes too expensive. However, since the fluctuations in the winding number result from the twisting and untwisting of the contour $D_k$, we expect that varying the number of steps $n$ should have a similar effect. Indeed, as shown in Fig.~\ref{fig:nstep}(b), after averaging over $n$, the root-mean-square $\delta\ze\nu\sim1/\sqrt{\Omega}$.

% Conclusion---
In conclusion, we have found universal fluctuations in the topological invariants characterizing a Floquet topological phase. We explained these fluctuations by positing a pseudorandom distribution of crossing points of the complex function whose winding number gives the topological invariant. This distribution follows from the diffusive process of emission, absorption, and motion of crossing points as frequency is lowered. Our results show that the limit $\Omega\to0$ has a rich structure that is distinct from the simple adiabatic limit: while the topological invariant vanishes~\cite{Note2} on average, consistent with the adiabatic limit, its fluctuations diverge. These fluctuation may be observed in the noise spectra of relevant quantities such as voltage noise~\cite{RodFerSer18a}, or by spectroscopic measures of the number of Floquet edge modes as recently observed in a photonic crystal emulator~\cite{ChePanWan18a}.

Universal fluctuations in Chern numbers have been studied in quantized classically-chaotic and random matrix theories~\cite{LebKurFei90a,WalWil95a}. By contrast, we study periodically driven systems, where topology is characterized not just by Chern numbers of a static Hamiltonian, but by independent winding numbers through a drive cycle. In this context, it would be interesting to study if driven systems with different symmetries (say, other than chiral symmetry) can support other universality classes of fluctuations of Floquet topological invariants. 

% Adiabatic limit

%acknowledgements
\begin{acknowledgements}
This work was supported in part by %Binational Science Foundation 
BSF grant No. 2014345, NSF CAREER grant DMR-1350663, and the College of Arts and Sciences at Indiana University. M.R.V. acknowledges the support and hospitality of Max Planck Institute for the Physics of Complex Systems.
\end{acknowledgements}

%\bibliographystyle{apsrev4-1}
%\bibliographystyle{h-physrev}
%\bibliographystyle{aipauth4-1}
%\bibliographystyle{physre}
%\bibliography{floquet,/Users/babak/LaTeXbibPath/MyBib.bib}

\onecolumngrid
%\appendix
\section{Supplemental Material }

In this Supplemental Material, we recap, for completeness, the derivation of the Floquet topological invariants in one-dimensional driven systems with a chiral reflection symmetry, and present the details of our numerical studies for driven Su-Schrieffer-Heeger (SSH), Kitaev, and SSH-Kitaev models, as well as the multi-step drive protocol in driven SSH model.

\subsection{Topological invariant for driven one-dimensional chiral-symmetric systems}
In this section, we briefly outline the derivation of the topological invariant
for one-dimensional chiral symmetric systems, following Ref. \cite{asboth2014}. 
A driven system has chiral symmetry if there exist a unitary, hermitian, local operator 
$\Gamma=\Gamma^\dagger=\Gamma^{-1}$, such that
\begin{equation}
\Gamma U(\tau,\tau+T) \Gamma = U^{-1}(\tau,\tau+T)
\label{eq:CS}
\end{equation}
where the time-ordered exponential $U(\tau,\tau+T) = \Texp\left[-i\hbar \int_\tau^{\tau+T} H(s)ds\right]$ is the full-period evolution operator with the initial time $\tau$ and period $T=2\pi/\Omega$ for the Hamiltonian $H(t)=H(t+T)$.
A topological invariant characterizing a chiral-symmetric one-dimensional system with perioidic boundary conditions can be written in the diagonal basis of $\Gamma$, where
%%%%%
\begin{equation}
H_{\text{eff}}(\tau) \equiv \frac iT \ln U(\tau,\tau+T) = \begin{bmatrix}
0       & h(\tau)   \\
h^\dagger(\tau) & 0 
\end{bmatrix},
\end{equation}
%%%%%
as
%%%%%
\begin{equation}
\nu_\tau=\mathcal{W}[h(\tau)]=\frac1{2 \pi i} \int_{-\pi}^{\pi} dk \frac{\partial}{\partial k} \ln [\mbox{det}\; h_k(\tau)].
\end{equation}
%%%%%
where $k$ is the lattice momentum.

Now, we say the drive protocol in $H(t)$ has chiral reflection symmetry if for some $\tau$ and $\tau' \in (0,T)$, one has $U(\tau+\tau',\tau+T)=\Gamma F^\dagger \Gamma$, where $F=U(\tau,\tau+\tau')$. That is, the drive protocol starting at $\tau$ has a reflection symmetry about $\tau+\tau'$.
Then, $U(\tau,\tau+T) = \Gamma F^\dagger \Gamma F $ and  $U(\tau+\tau',\tau+\tau'+T) = F \Gamma F^\dagger \Gamma$ both satisfy satisfy Eq. (\ref{eq:CS}). 

Thus, we can define two topological invariants $\nu=\nu_\tau$ and $\nu'=\nu_{\tau+\tau'}$. Physically, they 
can be interpreted as a bulk ``sublattice'' polarization at times $\tau$, and $\tau + \tau'$. Given
that states with quasienergy $\ze\epsilon=\Omega/2$ switch sublattice when they evolve from 
$\tau$ to $\tau + \tau'$, neither $\nu$ nor $\nu'$ alone are related to the number of edge states
in an open system, instead
\beq
\nu^\natural = \frac{1}{2}\left( \nu+\nu' \right), \;\;\; \nu^\flat = \frac{1}{2}\left( \nu-\nu' \right),
\enq
where $\nu^\natural$ and $\nu^\flat$ are the number of edge states with quasienergy $\zc\epsilon = 0$, and 
$\ze\epsilon = \Omega/2$, respectively. Using algebraic properties of winding numbers in the diagonal basis of $\Gamma$, the authors of Ref.~\cite{asboth2014} showed that $\nu^\natural$ and $\nu^\flat$ can be derived from the diagonal and off-diagonal blocks of $F = \begin{bmatrix}
A       & B   \\
C       & D 
\end{bmatrix}$ as $ \nu^\natural = \mathcal{W}[B]$, $ \nu^\flat = \mathcal{W}[D]$.
%
%
%%%%%%%%%%%%%%%%%%%%%%%%%%%%%%%%%%%%%%%%%%%%%%%%%%%%%%%%%%%%%%%%%%%%%%%%%%%%%%%%%%%%
\subsection{Driven SSH model}

The tight-binding Hamiltonian for driven the Su-Schrieffer-Heeger (SSH) model is given by the 
%%%%
\begin{equation}
\hat H (t) = \sum_r{[w + (-1)^r \delta(t)] \hat c^\dagger_{r+1} \hat c_r\nodag + \text{h.c.}}\;,
\end{equation}
%%%%
where $\hat c^\dagger_r$ creates a fermion at lattice site $r$, $w$ is the unmodulated hopping amplitude and $\delta(t)=\delta(t+2\pi/\Omega)$ is the hopping modulation, periodic in time with frequency $\Omega$.
For periodic boundary conditions, the crystal momentum 
$k$ is a good quantum number; defining the spinor $\hat c^\dagger_k = \sum_x e^{ik x } (\hat c^\dagger_{2x}, \hat c^\dagger_{2x+1})$, where $x=\floor{r/2}$ indexes the unit cells, we have $\hat H=\int  \hat c^\dagger_k h\nodag_k \hat c\nodag_k \frac{dk}{2\pi}$, with $k$ in the Brillouin zone $[-\pi,\pi]$, $h_k = d_{kx} \sigma_x + d_{ky} \sigma_y$ and
%%%%
%\begin{equation}
$
d_{kx} + i d_{ky} \equiv d_k = 2e^{ik/2} \left(w \cos \frac k2 + i\delta \sin \frac k2 \right).
$
%\end{equation}
%%%%
Discrete symmetries of this Hamiltonian include inversion $\sigma_x h_k \sigma_x = h_{-k}$, sublattice $\sigma_z h_k \sigma_z = -h_k$, and particle-hole symmetry $\sigma_yh_k^*\sigma_y = -h_k$, which place the static SSH model in the
BDI class  \cite{chingkai2016}. The instantaneous eigenvalues are $\pm |d_k|$ with $|d_k|=2\sqrt{w^2 \cos^2 \frac k2 + \delta^2 \sin^2 \frac k2}$.

\subsubsection{Quasienergy spectrum}

\begin{figure}[H]
	\centering
	\includegraphics[width=4.3in]{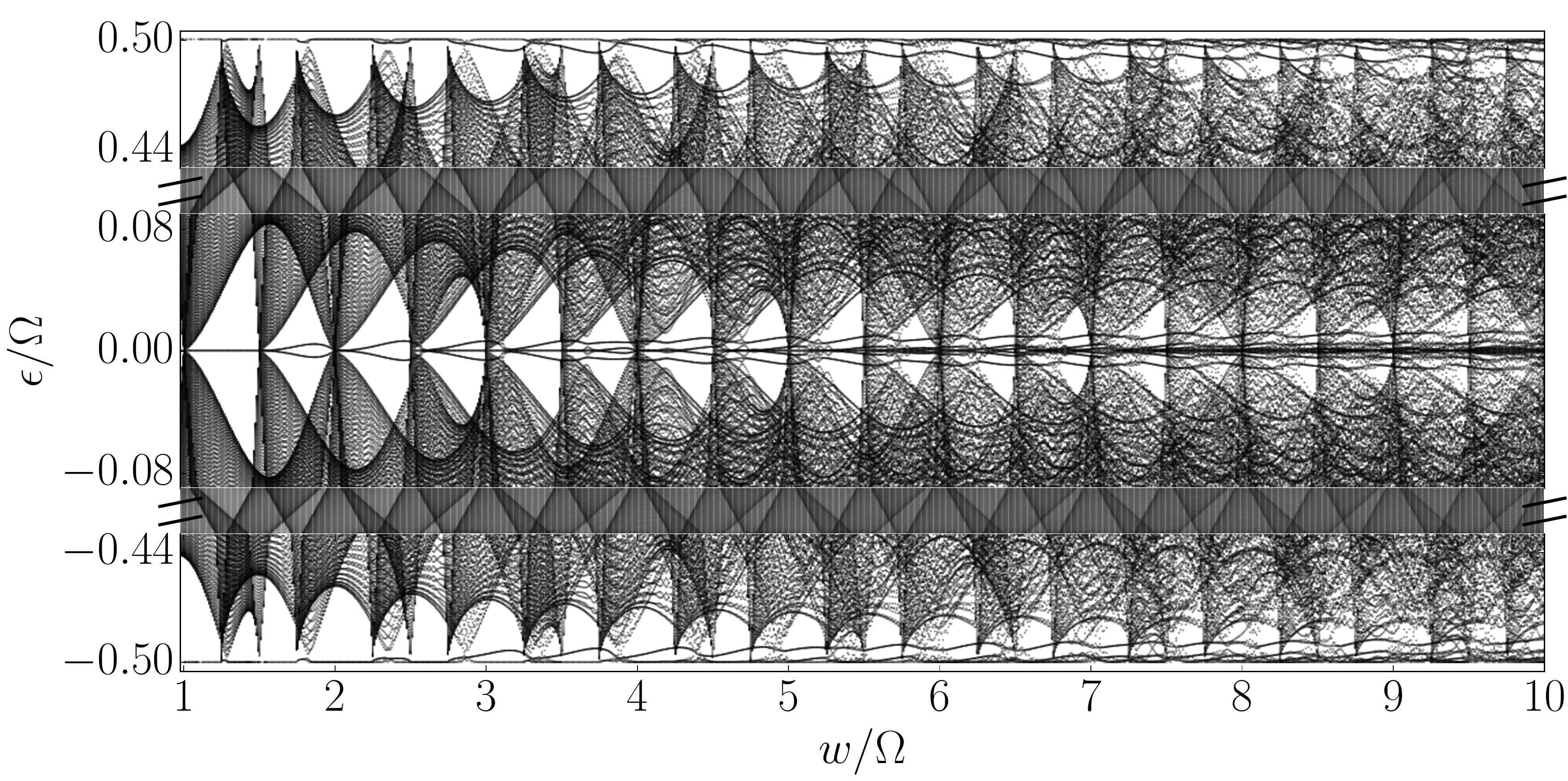}
	\caption{Quasienergy spectrum of the driven SSH model with open boundary conditions as a function of inverse frequency. The chain has $1000$ sites and is subject to a two-step drive protocol switching between $\delta_1/w = -0.75$ and $\delta_2/w = -0.25$. Edge states within quasienergy gaps at $\zc\epsilon=0$ and $\ze\epsilon=\Omega/2$ are found in an increasing number as the frequency is lowered.
	}
	\label{fig:spectrum_critical}
\end{figure}

\begin{figure}[H]
	\centering
	\includegraphics[width=16cm]{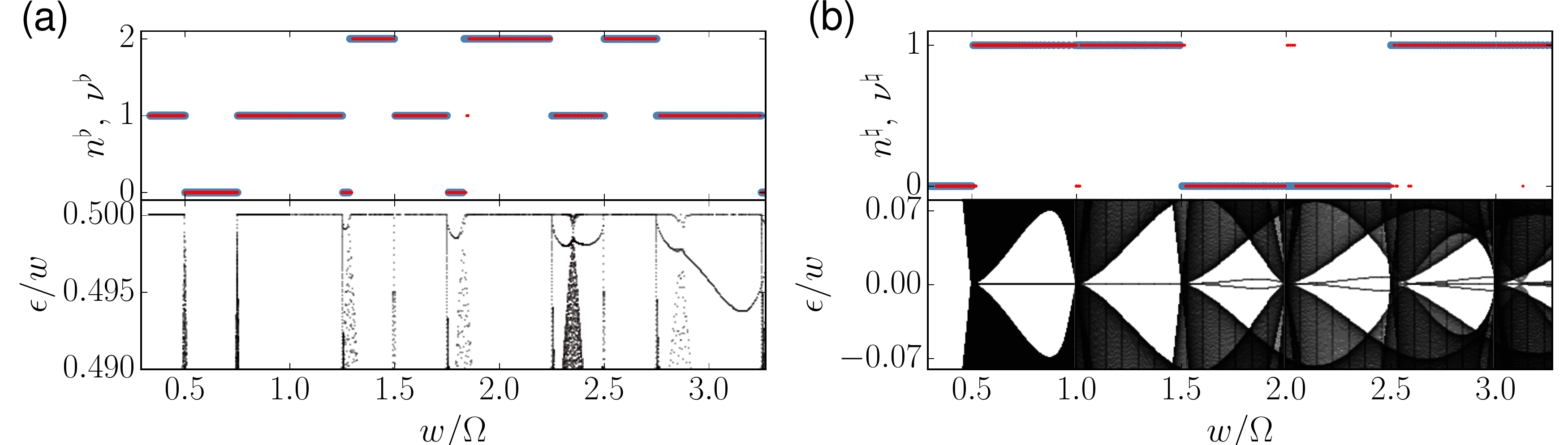}
	\caption{(a) Top: number of edge states $n^\flat$ (red)
		with quasienergy $\ze \epsilon=\Omega/2$ on top of
		the invariant $\nu^\flat$ (blue) as a function of inverse frequency. 
		Bottom: quasienergy spectrum around  $\ze\epsilon=\Omega/2$. 
		(b) Top: same as (a) for $n^\natural$ (red) and $\nu^\natural$ (blue). Bottom: quasienergy spectrum around $\zc\epsilon=0$.
		All the parameters are same as in Fig.~\ref{fig:spectrum_critical}.
	}
	\label{fig:spectrum_critical_2}
\end{figure}

In this section, we consider a driven SSH model with open boundary conditions,
and calculate the number of edge states present in the system at given frequency.
Fig.~\ref{fig:spectrum_critical} shows the quasienergies as a function of the
inverse frequency in the range $\Omega/w \in [0.1,1]$. The system has $N=1000$ sites and is driven by a two-step protocol
switching between $\delta_1/w = -0.75$ and $\delta_2/w = -0.25$. States with quasienergies
$\zc\epsilon=0$, and $\ze\epsilon=\Omega/2$ are localized at the edge of the 
system. 
Top panel of Fig.~\ref{fig:spectrum_critical_2}(a) [(b)] shows $n^\flat$ ($n^\natural$), the number of edge mode pairs with quasienergies $\ze\epsilon$ ($\zc\epsilon$), and are compared 
with the corresponding topological invariant $\nu^\flat$ ($\nu^\natural$). We use the same
parameters as in Fig.~\ref{fig:spectrum_critical}. In order to correctly resolve the
quasienergies, we consider a systems with $N=1000$ sites for the range $w/\Omega \in [0.3,1]$,
$N=2000$ for $w/\Omega \in [1,3]$, and  $N=3000$ for $w/\Omega \in [3,3.38]$. 

Both quantities are in agreement, as expected from the bulk-boundary correspondence. Deviations are observed at frequencies in proximity to gap closings, where
a larger system is required to correctly resolve the spectrum. Bottom panel 
of Fig.~\ref{fig:spectrum_critical_2}(a) [(b)]
shows a close-up in the quasienergy 
spectrum around $\ze\epsilon$ ($\zc\epsilon$). Changes in $\nu^\flat$ ($\nu^\natural$)
occur only when the gap around $\ze\epsilon$ ($\zc\epsilon$) closes. 

\subsubsection{Topological invariant $\nu^\natural$}

In this section, we discuss the low-frequency limit of $\nu^\natural$, the topological
invariant related to the number of edge states with quasienergy $\zc\epsilon$. 
As discussed in the main text, $\nu^\natural$ also presents fluctuations in the low-frequency
regime, with exceptions in special cases. For example, in the case of a symmetric drive, 
with $\delta_2=-\delta_1$ and $p_1=p_2$ we obtain 
$\zc\nu = 0$.

\begin{figure}[!h]
	\centering
	\includegraphics[width=6.0in]{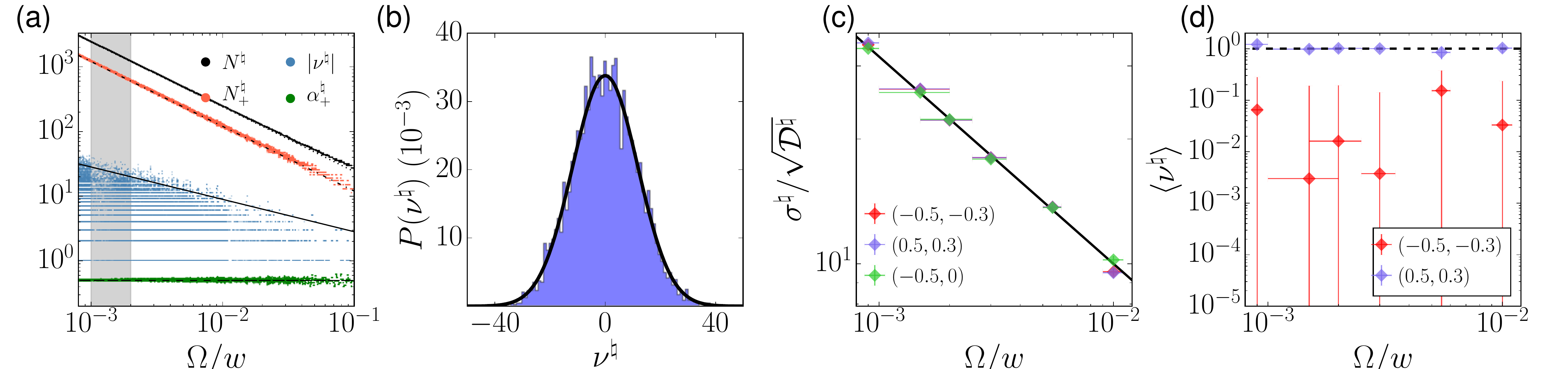}
	\caption{
		(a) Statistics of crossing points and winding number for a trivial protocol with $(\delta_1/w,\delta_2/w)=(-0.5,-0.3)$.
		From top to bottom: the total number of crossing points, $\zc N$, the positively-oriented crossing points $\zc N_+$, the ratio $\zc \alpha_+=\zc N_+/\zc N$, and the winding number $|\zc \nu|$ are calculated numerically at 30000 frequencies. The solid line shows $\sqrt{\zc{\mathcal{D}}/\Omega}$ with $\zc{\mathcal{D}} = ||\delta_1|-|\delta_2||$.
		(b) Probability distribution of $\zc\nu$ for the trivial protocol in (a) and frequency range $\Omega/w \in (1,2)\times10^{-3}$. The solid line is a Gaussian fit.
		(c) Standard deviation $\zc \sigma(\Omega)$ for various two-step drive protocols found by a Gaussian fit to $P(\zc\nu)$. The horizontal (vertical) line at each point indicates the range of frequencies (fitting error). The legend shows the values $(\delta_1/w, \delta_2/w)$. The solid line is $\zc \sigma = \sqrt{\zc{\mathcal{D}}/\Omega}$. (d) Average $\langle \zc \nu \rangle $, found by a Gaussian fit to $P(\zc\nu)$.}
	\label{fig:nuz_2}
\end{figure}

In general, we have to evaluate the invariant $\zc\nu$ numerically. Fig.~\ref{fig:nuz_2} 
summarizes our results. 
The winding number $\zc\nu$ fluctuates as frequency is lowered with a relative amplitude that grows as $\Omega\to0$.
Both $\zc N$ and $\zc N_+$ scale linearly with $1/\Omega$. The ratio $\zc \alpha_+=\zc N_+/\zc N$ approaches $1/2$ as the frequency decreases. Moreover, the range of $|\zc\nu|$ scales as $\sim1/\sqrt{\Omega}$. All of these properties are similar to the
low-frequency behavior of  $\ze\nu$. 
For a range of frequencies much lower than other energy scales in the system, we have determine the probability $P(\zc\nu)$ of finding $\zc\nu$ in our numerical histogram. As shown, $P(\zc \nu)$ follows a Gaussian distribution with a width that is given by $\sigma(\zc\nu) = \sqrt{\mathcal{D}^\natural/\Omega}$ over a decade of frequencies. When the
drive protocol is entirely in the static topological (trivial) phase, the probability $P(\zc\nu)$
is centered at  $\zc\nu = 1$ ($\zc\nu=0$).

\subsubsection{Multi-step drive protocol}
In this section, we consider a multi-step driving protocol as an approximation to
the smooth protocol $\delta(t) = \bar \delta + A \cos(\Omega t)$. For a protocol with $n$ steps, the
half-propagator is given by
\beq
U_k(\pi/\Omega) = \Pi_{j=0}^{n} e^{-i h_k(j \pi /(\Omega n)) \pi/(\Omega n)}\; .
\enq
In the limit $n \rightarrow \infty$ we recover the smooth protocol. In Fig. \ref{fig:nstep}(a)
we plot the invariant $|\ze \nu|$ and the ratio $\ze \alpha_+$
as a function of the frequency $\Omega$ for a protocol with $n=20$ steps. As in the case of a two-step drive,
we find that the ration $\ze \alpha_+$ approaches $1/2$ as the frequency decreases, suggesting a
diffusive process. Also, $|\ze \nu|$ presents fluctuations that grow as $\sim 1/\sqrt{\Omega}$. 
For large $n$, it becomes difficult to accurately calculate the invariant, and 
collecting enough statistics as a function of frequency to determine the scaling in the low-frequency
limit becomes more challenging. However, varying $n$ changes the structure of complex function $D_k$, 
creating new twists. Therefore, we compute $\ze \nu$ for different frequencies $\Omega$
as a function of $n$ in the range $[2,50]$ and consider the average over $n$. In Fig.~\ref{fig:nstep}(b), we plot the root-mean-square
$\delta \ze \nu$ as a function of frequency and find that it also scales as $1/\sqrt{\Omega}$. 
%
%
%%%%% Fig. 4 %%%%%%
\begin{figure}[h]
	\centering
	\includegraphics[width=16.0cm]{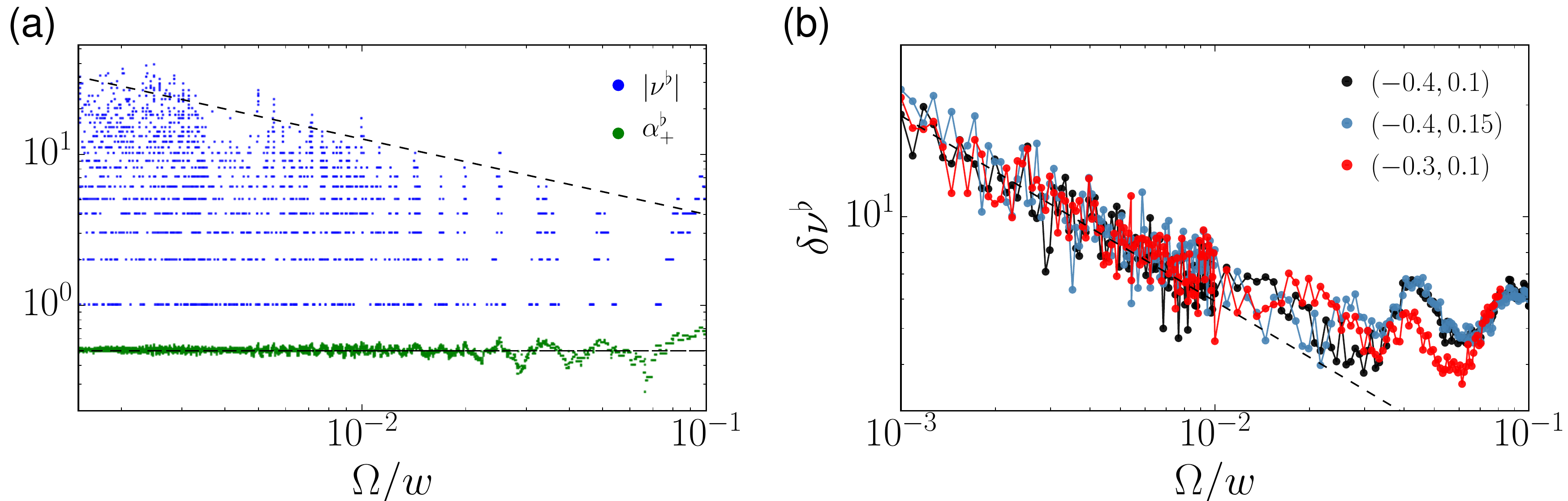}
	\caption{(a) Winding number, $|\ze\nu|$, and the ratio $\ze\alpha_+$ 
		in a multi-step protocol as a function of the frequency $\Omega$. The number of steps in the drive is $n=20$
		with the mean value $\bar\delta/w =-0.4$ and amplitude $A/w = 0.1$. From top to bottom, the dashed lines
		correspond to $\sqrt{0.35/\Omega}$ and $1/2$.
		(b) The root-mean-square $\delta\ze\nu$, averaged over $n$ as a function frequency for three sets of $(A,\bar \delta)$.
		The dashed line corresponds to $\sqrt{\ze{\mathcal{D}}/\Omega}$, fitted with $\ze{\mathcal{D}}/w = 0.5 $.	}
	\label{fig:nstep}
\end{figure}
%%%%%%%%%%%%%%%

%
\subsection{Driven Kitaev Model}
In this section we consider the driven Kitaev model, and study the behaviour of the topological
invariants in the low-frequency regime. The Hamiltonian  is given by \cite{kitaev2001}
\beq
\hat H = \sum_{r=1}^{N-1} \left( w \hat c^\dagger_{r+1} \hat c\nodag_{r} + \Delta \hat c^\dagger_{r+1} \hat c^\dagger_{r} + h.c.\right) -\mu \sum_{r=1}^N \hat c^\dagger_r \hat c\nodag_{r},
\enq
where $w$ and $\Delta$ are the nearest-neighbor hopping
and pairing amplitudes respectively, and $\mu$ is the
chemical potential. Imposing periodic boundary conditions, and introducing the Nambu
spinors $\hat \psi_k^\trans = \left(\hat c\nodag_k, \hat c^\dagger_{-k} \right)$, the Hamiltonian can be written as 
$\hat H = \frac12 \sum_k \hat \psi^\dagger_k h\nodag_k \hat \psi\nodag_k - \mu N$, where 
$
h_k = \left( 2 w \cos k - \mu  \right) \sigma_z + 2 |\Delta| \sin k \left( \cos\phi\:  \sigma_y
-\sin\phi\: \sigma_x \right).
$
This has the particle-hole symmetry $h_k = - \sigma_x h_{-k}^*\sigma_x$. %From now on, we fix the order parameter phase $\phi=\pi$.
Upon the rotation $R = \exp( - i \sigma_y \pi/4)\exp(i\phi\sigma_z/2)$, the Hamiltonian is transformed to
\beq
h_k \to R h_k R^\dagger = \left( 2 w \cos k - \mu  \right) \sigma_x  + |\Delta| \sin k\:  \sigma_y,
\enq
as shown in the main text. In this representation, the Kitaev model has the discrete symmetries $h_k = - \sigma_z h^*_{-k} \sigma_z$, and
$h_k = h^*_{-k}$. %, and $ \sigma_z h_k \sigma_z = -h_k$. \babak{This is not independent from the other two.}
The instantaneous eigenvalues are  
$\pm \sqrt{\left( 2w \cos k - \mu  \right)^2
	+|\Delta|^2 \sin^2 k}$.

In a system with open boundary conditions we
find one pair of Majorana states at zero energy for $|\mu| < 2 |w|$, and zero for $|\mu| > 2|w|$.
At $|\mu| = 2 w$ the gap closes. The static topological invariant gives $\nu=1$ for $|\mu| < 2 w$
and zero otherwise.
\begin{figure}[H]
	\centering
	\includegraphics[width=15.5cm]{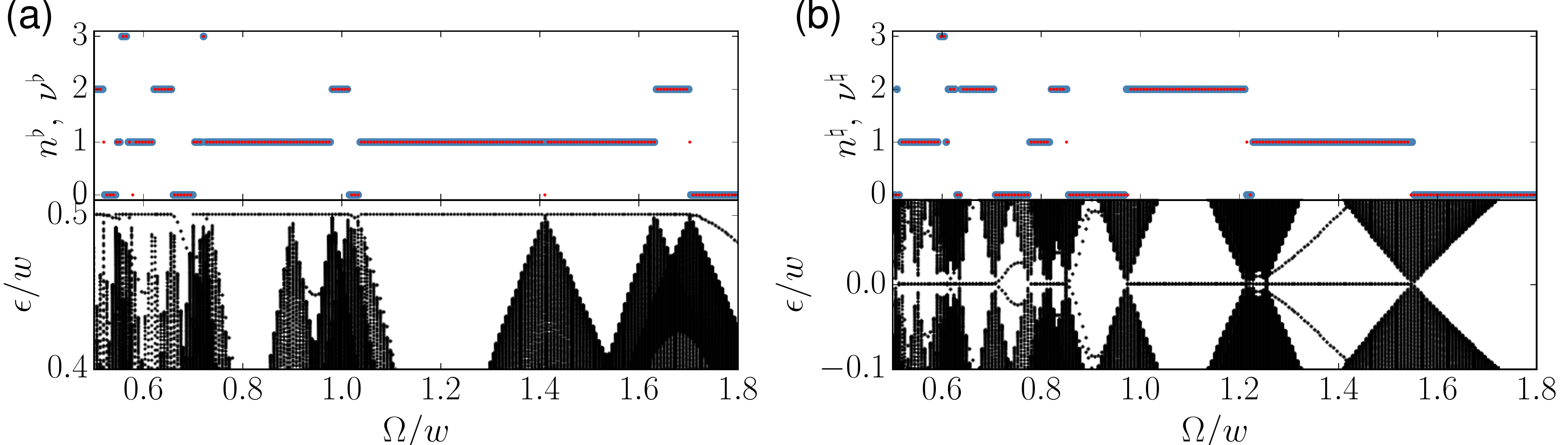}
	\caption{(a) Top: number of edge states $n^\flat$ (red)
		with quasienergy $\ze \epsilon=\Omega/2$ on top of
		the invariant $\nu^\flat$ (blue) as a function of  frequency. 
		Bottom: quasienergy spectrum around  $\ze\epsilon=\Omega/2$ for a system with open
		boundary conditions and $N=1000$ sites. 
		(b) Top: same as (a) for $n^\natural$ (red) and $\nu^\natural$ (blue). Bottom: quasienergy spectrum around $\zc\epsilon=0$.
		The driving protocol is $(\mu_1/w=1.0, \mu_2/w=-1.9)$, and $|\Delta|/w=1$.
	}
	\label{fig:kitaev_1}
\end{figure}

Choosing a suitable driving protocol, 
the driven Kitaev model also has chiral symmetry, and the topological invariants are given 
by $\ze \nu$ and $\zc \nu$, as for the SSH model.
In this work, we choose to drive the chemical potential $\mu(t)$ with the two-step protocol
$\mu(t) = \mu_1$ for $0<t<2\pi p_1/\Omega$ and $\mu_2$ for $2\pi p_1/\Omega<t<2\pi/\Omega$, where $0<p_1<1$. We fix $p_1=0.5$. The bottom panels in figure~\ref{fig:kitaev_1}
show the quasienergy spectrum for a chain with
$N=1000$ sites in the frequency range $\Omega/w \in [0.5,1.8]$ around the Floquet zone edge and center.
Top panel (a) [(b)] shows the topological invariant $\ze \nu$ ($\zc\nu$) underneath the number of states with quasienergy
$\ze \epsilon$ ($\zc \epsilon$), shown in red. These quantities are in agreement. Differences can be found at frequencies where the corresponding quasienergy gap is closed.     

\begin{figure}[!h]
	\centering
	\includegraphics[width=14.0cm]{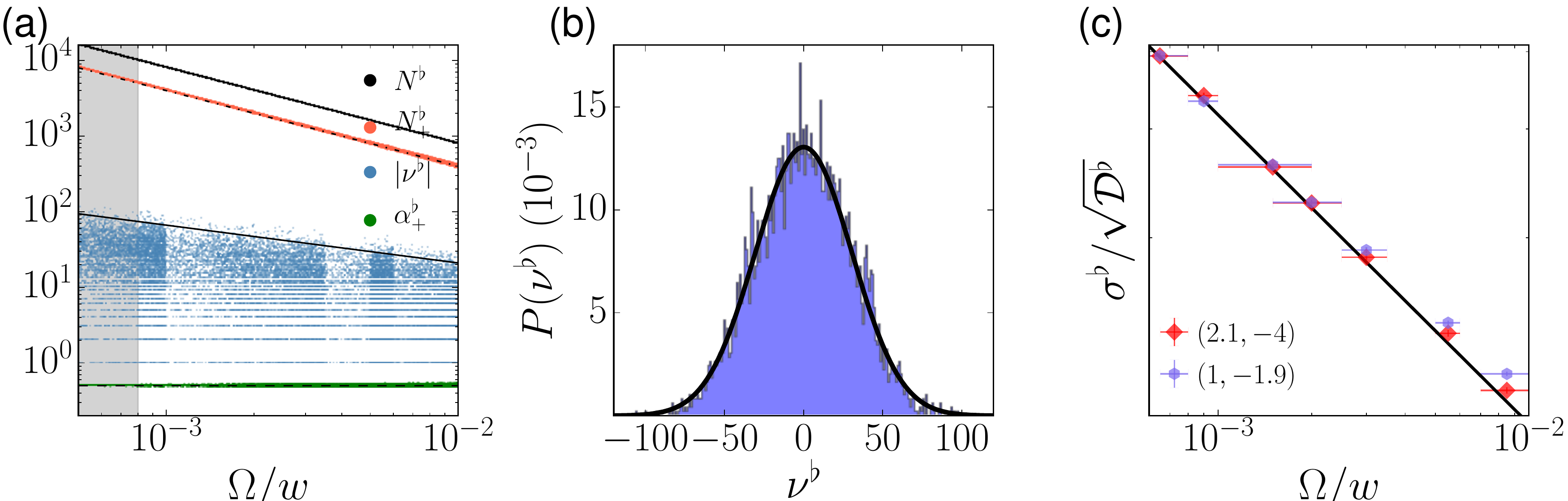}
	\caption{Topological invariant for a driven Kitaev chain. (a) $N^\flat$, $N^\flat_+$, $\alpha^\flat_+ = \ze N_+/\ze N$, and $\nu^\flat$ as a function of the frequency for protocol alternating the chemical potential between
		$\mu_1/w=2.1$ and $\mu_2/w=-4$. In panel (b) we plot the probability to obtain $\nu^\flat$ in the range of frequencies shown shaded in (a). (c) shows the width of the Gaussian distribution as a function of frequency for two sets of chemical potentials $(\mu_1/w, \mu_2/w)$. The diffusions constants are $D^\flat=0.81$  and $D^\flat=0.62$ for $(2.1,-4)$, and $(1,-1.9)$ respectively. The vertical lines indicate the range of frequencies used to calculate the distribution, and the horizontal lines indicate the fitting error. The superconducting gap is fixed in the protocol at $|\Delta|/w=1$. For each point shown, we considered $5000$ frequencies.
	}
	\label{fig:kitaev_2}
\end{figure}

We now evaluate the topological invariants numerically in the low-frequency regime. For concreteness, we
focus on $\ze \nu$. Our results are summarized in Fig.~\ref{fig:kitaev_2}. 
As for the SSH model, we find that the 
range of $|\ze\nu|$ scales as $\sim 1/\sqrt{\Omega}$,
both $\ze N$ and $\ze N_+$ scale linearly with $1/\Omega$, and the ratio 
$\zc \alpha_+$ approaches $1/2$ as the frequency decreases. Furthermore, the
probability $P(\ze \nu)$ follows a Gaussian distribution with width $\sigma(\ze\nu) = \sqrt{\mathcal{D}^\flat/\Omega}$. 

\subsection{Four-band driven Kitaev-SSH Model}
In this section we consider the Kitaev-SSH model \cite{wakatsuki2014}. The Hamiltonian is given by 
\begin{align} 
	H = - \sum_x &\left\{ w \left[(1+\delta/w) \hat c^\dagger_{B,x} \hat c\nodag_{A,x}
	+ (1-\delta/w) \hat c^\dagger_{A,x+1} \hat c\nodag_{B,x} \right] \right. \nonumber\\
	&\hspace{-1mm}\left. -\Delta \left[(1+\delta/w) \hat c^\dagger_{B,x} \hat c^\dagger_{A,x} + (1-\delta/w) \hat c^\dagger_{A,x+1} \hat c^\dagger_{B,x}\right]
	- \frac\mu2 \left( \hat c^\dagger_{A,x}\hat c\nodag_{A,x}+
	\hat c^\dagger_{B,x} \hat c\nodag_{B,x} \right) \right\} + \text{h.c.},
\end{align}
where $x$ labels the unit cells, and $A$, $B$ label the sublattices in each unit cell, $w+\delta$ ($w-\delta$) 
and $\Delta$ are the intra-cell (inter-cell) hopping
and pairing amplitudes respectively and $\mu$ is the
chemical potential. 
If we consider periodic boundary conditions, and define the spinor 
$ \hat \psi^\dagger_k = \left(\hat c^\dagger_{kA}, \hat c^\dagger_{kB}  , \hat c\nodag_{-kA}, \hat c\nodag_{-kB} \right)$, the momentum-dependent Hamiltonian takes the form \cite{wakatsuki2014}
\beq
h_k = \begin{bmatrix}
	-\mu & z_k & 0 & \Delta_k   \\
	z_k^*  & -\mu & -\Delta_k^* & 0 \\
	0    & -\Delta_k & \mu & -z_k \\
	\Delta_k^* & 0 & -z_k^* & \mu
\end{bmatrix},
\enq
where $z_k=-\left[ (w+\delta)+(w-\delta)e^{-ik} \right]$,
and $\Delta_k=-\Delta \left[ (1+\delta/w)-(1-\delta/w)e^{-ik} \right]$. For our purpose, the most important feature of this model is that it has four energy bands, given by
$
\pm \sqrt{  \mu^2 + |z_k|^2 + |\Delta_k|^2 \pm 2\sqrt{\mu^2 |z_k|^2+(4 \Delta \delta )^2 } }
$

For $\delta/w<-0.2 $ we have two pairs of zero-energy states
localized at the edge, while for $-0.2<\delta/w<0.2 $ we have only one pair. 
For $\delta/w>0.2 $ there are no zero-energy states.
For simplicity, now on we consider the case
$\mu_1/w = \mu_2/w = 0$, although is possible to also consider finite chemical potentials. 
The chiral basis for this model is 
defined by the transformation 
$
\Gamma = \frac12(\tau_0 \otimes \sigma_0 + 
\tau_\alpha \otimes \sigma_\alpha),
$
where $\sigma_\alpha$ and $\tau_\alpha$ for $\alpha=x,y,z$
are the Pauli matrices
in particle-hole and sublattice space, and
$\tau_0=\sigma_0=\id$.
%
%\beq
%\Gamma = \begin{bmatrix}
%	1 & 0 & 0 & 0 \\
%    0 & 0 & 1 & 0 \\
%  	0 & 1 & 0 & 0 \\
%	0 & 0 & 0 & 1   	
%\end{bmatrix}.
%\enq

\begin{figure}[!h]
	\centering
	\includegraphics[width=16.0cm]{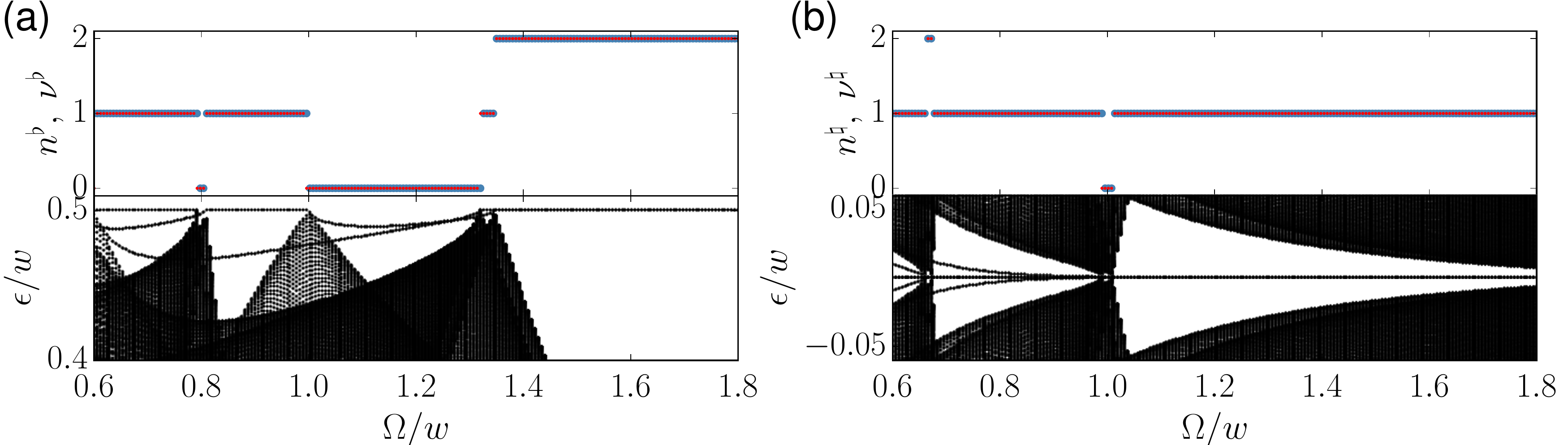}
	\caption{Top panel: (a) [(b)] Number of edge states  $\ze n$ ($\zc n$) with quasienergy $\ze \epsilon$ 
		($\zc \epsilon$), and topological
		invariant $\ze \nu$ ($\zc \nu$) as a function of frequency. Bottom panel:
		Quasienergy spectrum as a function of $\Omega$ for a driven Kitaev-SSH chain with open boundary conditions. The
		system has $N=1000$ sites, $\Delta/w=0.2$, and the driving protocol $(\delta_1/w=3.0, \delta_2/w=-2.0)$.
	}
	\label{fig:kitaev_3}
\end{figure}

We choose the driving protocol  $\delta(t) = \delta_1$ for $0<t<2\pi p_1/\Omega$ and $\delta_2$ for $2\pi 	p_1/\Omega<t<2\pi/\Omega$, where $0<p_1<1$ to ensure that the driven system posses chiral symmetry.
As before, we will fix $p_1=0.5$. 
In Fig.~\ref{fig:kitaev_3}, the bottom panels show the quasienergy spectrum for a driven chain 
with $N=1000$ sites around the Floquet zone edge (a) and center (b). The rest of the 
parameters used are  $\delta_1/w=3.0$, $\delta_2/w=-2.0$, and $|\Delta|/w=0.2$. 
Top panel (a) [(b)] shows the topological invariant $\ze \nu$ ($\zc \nu$) underneath 
of $\ze n$ ($\zc n$), the number of states with quasienergy $\ze \epsilon=\Omega/2$ ($\zc \epsilon=0$), shown in red. These quantities are in agreement, as expected from the bulk-edge correspondence. 

\begin{figure}[h]
	\centering
	\includegraphics[width=15.0cm]{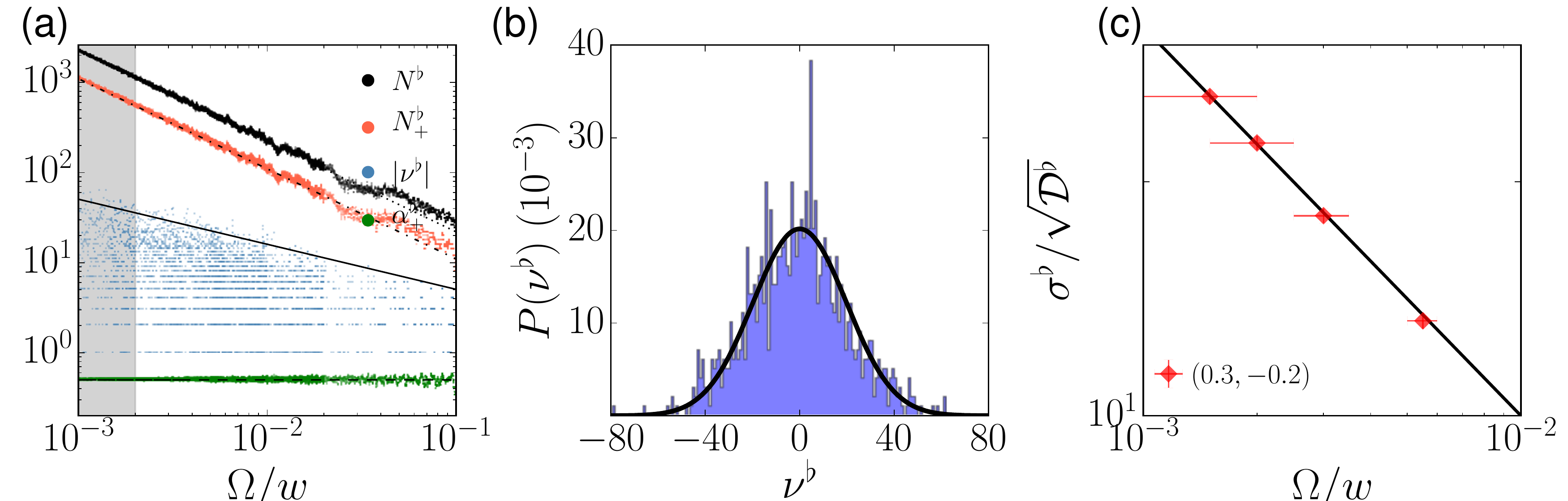}
	\caption{Topological invariant for a driven Kitaev-SSH chain. (a) $N^\flat$, $N^\flat_+$, $\alpha^\flat_+ = N_+/N$, and $\nu^\flat$ as a function of the frequency. In panel (b) we plot the probability to obtain $\nu^\flat$ in the range of frequencies shown shaded un subpanel (b). Panel (c) shows the width of the Gaussian distribution as a function of frequency for $(\delta_1=0.3, \delta_2=-0.2)$. The diffusions constant is $D^\flat=0.59$. The vertical lines indicate the range of frequencies used to calculate the distribution, and the horizontal lines indicate the fitting error. The superconducting gap is fixed in the protocol at $|\Delta|/w=0.2$. For each point shown, we considered $1000$ frequencies.
	}
	\label{fig:kitaev_4}
\end{figure}

The topological invariants are derived from the half-period evolution propagator
$
U_k(\pi/\Omega) \equiv \left(\begin{array}{cc} A_k & B_k \\ C_k & D_k  \end{array}\right),
$
where $A_k,B_k,C_k$ and $D_k$ are now $2\times2$ matrices. The winding numbers are given by $\nu^\flat = \mathcal{W}[D]$, and $\nu^\natural = \mathcal{W}[B]$
where 
$
\mathcal{W}[h]=\frac1{2\pi i} \int_{-\pi}^\pi \frac{\partial}{\partial k}
\ln \left( \mbox{det} \; h_k \right)dk.
$
As explained in the main text, 
for multi-band systems, the invariant
is still defined by the winding of a complex functions, $\det D_k$ and $\det B_k$. The evaluation of $\nu^\flat$ is done numerically,
and our results are summarized in Fig.~\ref{fig:kitaev_4}. As in the other two driven chiral-symmetric systems,
the SSH model and the Kitaev model, we find that $\ze\nu$ fluctuates as frequency is lowered.
The probability $P(\ze \nu)$ also follows a Gaussian distribution, with a width given by $\sigma(\ze\nu) = \sqrt{\mathcal{D}^\flat/\Omega}$.

\end{document}